\documentclass{JHEP3}

\usepackage{graphicx}
\usepackage{mathrsfs}
\usepackage{amsmath, amsthm, amssymb}

\newcommand{\be}{\begin{equation}}
\newcommand{\ee}{\end{equation}}
\newcommand{\beq}{\begin{equation}}
\newcommand{\eeq}{\end{equation}}
\newcommand{\bea}{\begin{eqnarray}}
\newcommand{\eea}{\end{eqnarray}}

\newcommand{\gsim}{\lower.7ex\hbox{$\;\stackrel{\textstyle>}{\sim}\;$}}
\newcommand{\lsim}{\lower.7ex\hbox{$\;\stackrel{\textstyle<}{\sim}\;$}}

\title{Understanding and improving the Effective Mass for LHC searches}

\author{Maria Eugenia Cabrera\\
        University of Amsterdam\\
        Institute of Theoretical Physics\\
        GRAPPA\\
        E-mail: \email{M.E.CabreraCatalan@uva.nl}}

\author{J. Alberto Casas \\
        Instituto de F\'isica Te\'orica, IFT-UAM/CSIC \\
        U.A.M., Cantoblanco, \\
        28049 Madrid, Spain \\
        E-mail: \email{alberto.casas@uam.es}} 

\abstract{\small
A handy and extensively used kinematic variable in LHC analyses, especially for golden SUSY signals of multijets plus missing energy, is the Effective Mass, $M_{\rm eff}=\sum_{j}
|{p}_T^{\ j}|+ |{p}_T^{\ \mathrm{miss}}|$. Empirically, the value of $M_{\rm eff}$ at which the histogram of events has a maximum is correlated with the SUSY spectrum, $\left. M_{\rm eff}\right|_{\rm max}\simeq 80\%\ M_{\rm susy}$, where $M_{\rm susy}$ is essentially the sum of the masses of the SUSY particles initially created. In this paper we explain the reason for such strong correlation, pointing out the cases where the correlation is not good. Besides, we propose a new variable, the Effective Transverse Energy, ${\cal E}_T^{\rm eff}$, which shows an even better and more direct correlation $\left. {\cal E}_T^{\rm eff}\right|_{\rm max}\simeq M_{\rm susy}$, and is independent of the procedure followed to identify the jets.
${\cal E}_T^{\rm eff}$ and $M_{\rm eff}$ are complementary variables, rather than competitors; and plotting histograms in both can be useful to cross-check the results, allowing a better and more robust identification of $M_{\rm susy}$. The extension of this procedure to other scenarios of new physics (not necessarily SUSY) is straightforward.
}

\preprint{IFT-UAM/CSIC-12-64}

\begin{document}

\section{Introduction}

The LHC is already probing new physics beyond the reach of past
experiments. There are two main questions to address: 1) Is there any
signal of New Physics (NP)? and 2) In the positive case, which NP is
it? In order to optimize the answer to these questions there is an
intense activity to explore assorted strategies for the search of
NP. The task is challenging, due in part to the fact that LHC data,
though very rich, are not as clean as those from an $e^+\,e^-$
collider. 

Supersymmetry (SUSY) is one of the few candidates for physics beyond the SM
that is really well motivated from the theoretical point of view and
allows to perform detailed calculations and thus realize precise
predictions. Indeed SUSY has been the most extensively studied
candidate for new physics in the last decades and the first LHC
analyses are using SUSY as a paradigmatic scenario of new physics to
present their constraints on physics beyond the SM. Of course, this
does not mean that SUSY is really there, but clearly is a most serious
scenario to be considered in the light of the LHC.

In the framework of SUSY and under the assumption that superparticles are light
enough to be produced and detected at LHC, one can expect a typical
signal of multijets and large missing energy with or without leptons.
Searches using channels with leptons in the final state allow quite clean
reconstructions. Looking for end-points and using ingenious kinematic
variables \cite{Lester:1999tx,Barr:2003rg,Lester:2007fq,Cho:2007qv,Barr:2007hy,
Burns:2008va,Randall:2008rw,Polesello:2009rn,Cho:2009ve,Kim:2009si,Konar:2009wn},
one can determine in some cases the mass of the sparticle produced in the
decay chain.
On the other hand, the channel with zero leptons tends to be the golden one
producing signals beyond the Standard Model. However, here the analysis
becomes much more complicated and the correct identification and
reconstruction of the properties of the final states is quite tough. Besides,
the separation of the decay chains is normally not possible.

In this context, the use of appropriate kinematic variables and the choice of
optimal cuts are instrumental to maximize the SUSY signal from the Standard
Model background. But, for the multijet channel, finding the optimal choice is
a difficult task. In ATLAS analyses the Effective Mass variable ($M_{\rm
  eff}$) plays a very important role \cite{Aad:2009wy,daCosta:2011qk,Aad:2011ib}. It is defined as,
\begin{eqnarray}
\label{eq:meff4}
M_{\rm eff}^{(4)}=\sum_{j=1}^4
|{p}_T^{\ j}|+|{p}_T^{\ \mathrm{miss}}|,
\end{eqnarray}
where ${p}_T^{\ j}$ is the transverse component of the momentum of the four hardest
jets [thus the superscript (4)] and ${p}_T^{\ \mathrm{miss}}$ is the missing transverse
momentum. 

It was found in \cite{Hinchliffe:1996iu} (from now on HPSSY) that for the Constrained Minimal Supersymmetric Standard Model (CMSSM) there is an outstanding correlation between the value of $M_{\rm eff}^{(4)}$ at which the histogram of events has a maximum and the
supersymmetric mass $M_{{\rm susy}}$ defined as the minimum of the squark and gluino masses,
\begin{eqnarray}
\label{eq:Msusy1}
M_{\rm susy}^{\rm (HPSSY)}={\rm min}\{m_{\tilde{q}},m_{\tilde{g}}\}.
\end{eqnarray}

More precisely, HPSSY showed that systematically $M_{\rm eff}^{(4)}|_{\rm max}\simeq 1.9\ M_{\rm susy}^{\rm (HPSSY)}$ for a large collection of CMSSM models with assorted values of the initial supersymmetric parameters. This is very remarkable. However there are some caveats:

\begin{enumerate}

\item As admitted by HPSSY the definition of the supersymmetric mass as in eq.(\ref{eq:Msusy1}) was rather arbitrary. Although typically the production of a pair of the lightest supersymmetric particles (among squarks and gluinos) is favoured, the production of other pairs, like squark-gluino, can be very frequent too. A refined definition of $M_{\rm susy}$ was proposed in \cite{Tovey:2000wk}, namely the sum of all supersymmetric masses weighted by the production cross section (normalized to one) of each one. In this way, the correlation behaves even better, though since the new $M_{\rm susy}$ is slightly larger the above number $1.9$ decreases somewhat.

\item On the other hand, $M_{\rm eff}^{(4)}$ is also smaller than the total effective mass, $M_{\rm eff}$ . A more compelling definition, from the theoretical perspective, would be to sum in eq.(\ref{eq:meff4}) over all jets (surviving certain cuts); though this might be problematic from a practical point of view.

\item So far there is no theoretical explanation for the strong correlation found between the effective mass and the supersymmetric mass. In ref.\cite{Hinchliffe:1996iu} it is presented as an empirical fact. Certainly, for events with large transverse momenta, one would expect the effective mass to be of the order of sum of the masses of the supersymmetric particles initially created. However, it is not clear why the maximum number of events is always reached at a value of the effective mass so strongly correlated (but slightly smaller) to that sum.

\item As a matter of fact, it was shown in \cite{Tovey:2000wk} that for more generic MSSM models (beyond the CMSSM) such correlation fails in many instances. Again, it would be very useful to know the reason for that, in order to avoid flawed analyses and improve general strategies.  

\end{enumerate}

In the present paper we will offer an explanation for the above points (3) and (4). This understanding will allow us to propose a new kinematic variable, alternative to the effective mass, which shows an even better correlation with the supersymmetric masses, and can be used as an alternative or complementary check in multijet studies at LHC. The discussion and proposal presented here can be easily extrapolated to other scenarios of New Physics different from SUSY.

In section 2 we introduce some relevant kinematic concepts and establish an
  analogy between the production of the W boson and that of Supersymmetric
  particles in hadron colliders. In section 3 we explain the empirical strong correlation between $M_{\rm eff}$ and
  $M_{\rm susy}$. In section 4 we propose a new kinematic variable, ${\cal E}_T^{\rm eff}$, which shows
  an even more robust correlation with $M_{\rm susy}$. In section 5 we test the
  efficiency of $M_{\rm eff}$ and ${\cal E}_T^{\rm eff}$, and compare their
  behaviour using signal simulations at the LHC.


%
%
\section{Strategy for the kinematic analysis}

\subsection{W boson production in $p\,\bar{p}$ colliders }

We start by briefly reviewing the $W$ boson production and decay in a $p\,\bar{p}$ collider,  since it is the
simplest case involving visible and invisible particles, and it is therefore a useful guide to introduce some notation and some relevant kinematic concepts which will be used later on.

In a $p\,\bar{p}$ collider the main process for $W$ boson production
is through quark-antiquark annihilation. Of course, the laboratory (LAB) system of reference does not coincide in general with the center-of-mass (CM) one, since it is affected by boosts, mainly along the collision line ($z$). In addition, there can be less important boosts along a transverse direction due to other effects; in particular, the quarks in initial state
may radiate soft gluons (some of them energetic enough to be detected
as jets), but for the moment we will ignore these effects. 
\begin{figure}[t]
\begin{center}
\includegraphics[width=0.35\linewidth]{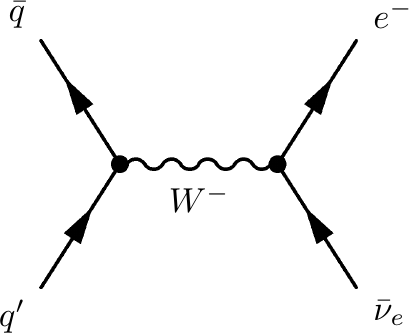}
\end{center}
\caption{Feynman diagram of W-boson production and decay at hadron colliders.}
\label{fig:Wprod}
\end{figure} 

The $W$ boson mass is measured by analyzing its leptonic decay
$q\bar{q}\rightarrow W\rightarrow e\bar\nu$, see Fig.\ref{fig:Wprod}. From the energy and momentum
of the electron and the (anti)neutrino, one can derive the invariant mass of the
$W$ boson,
\begin{eqnarray}
\label{eq:Minv_W}
M_W^2=M_{\rm inv}^2 &=& (E^e + E^\nu)^2 - (\vec{p}^{\ e} +
\vec{p}^{\ \nu})^2\\ \nonumber &=& m_e^2 + m_\nu^2 + 2\, [\,E^e_T
  \,E^\nu_T\cosh (y^e-y^\nu) - \vec{p}^{\ e}_{T}\cdot
  \vec{p}^{\ \nu}_{T}\,]\ ,
\end{eqnarray}
where $p_T^a$ is the transverse component of the 3-momentum of the $a-$particle, while $E_T^a$ and $y^a$ are respectively its transverse energy and rapidity, defined as 
\begin{eqnarray}
\label{eq:ET:Y}
(E_T^a)^2 = m_a^2 + ({p}_T^a)^{\ 2} = (E^a)^2 - (p_z^a)^2 \;\; ,\qquad y^a =
\frac{1}{2}\ln \left( \frac{E^a+p_z^a}{E^a-p_z^a} \right)\ .
\end{eqnarray}
Note that $E_T^a$ is invariant under boosts in the $z-$direction. $E^a$ and $p_Z^a$ can be written in terms of $E_T^a$ and $y^a$ as
\begin{eqnarray}
\label{eq:E:Pz}
E^a = E_T^a \cosh{y^a} \; ,\qquad p_z^a = E_T^a \sinh{y^a}.
\end{eqnarray}

Since it is not possible to measure the neutrino momentum, one cannot
directly determine the $W$ invariant-mass from the experiment using eq.(\ref{eq:Minv_W}). Instead, one can consider the differential cross section of production of an electron-neutrino pair. 

Let us start working in the CM system. 
Calling $\theta, \phi$ the polar and azimuthal angles of the electron 3-momentum, the corresponding differential phase-space factor is proportional to the solid angle element $d \sigma \propto d^2\Omega = d \cos \theta d \phi$, where the integration over $\phi$ can be performed in a trivial way using symmetry around the $z-$axis. Now, one can consider the distribution of cross section for different values of $p^e_T$, i.e.
\begin{eqnarray}
\label{dsigma}
d \sigma \propto \frac{d \cos \theta}{d p_T^e} d p_T^e\ ,
\end{eqnarray}
where the Jacobian factor reads
\begin{eqnarray}
\label{Jacobian}
\frac{d \cos \theta}{d p_T^e} &=& \frac{-p_T^e |\vec{p_e}| }{\sqrt{|\vec{p_e}|^2-(p_T^e)^2}}
\nonumber\\
&=& \frac{-2E\ p_T^e |\vec{p_e}| }{ \sqrt{E^2- (E_T^e + E_T^\nu)^2}\sqrt{E^2- (E_T^e - E_T^\nu)^2}}\ ,
\end{eqnarray}
with $E=E^e+E^\nu$ denoting the total energy at CM ($E=M_W$ if the $W$ is produced on-shell). Of course the neutrino and electron masses are completely negligible in this context, but for future convenience we will maintain them in the expressions.

The phase-space factor (\ref{Jacobian}) shows a {\em pole} at $E_T^e + E_T^\nu = E$. Denoting
\begin{eqnarray}
\label{ETW}
{\cal E}_T =E_T^e + E_T^\nu \ =\ \sqrt{(p_T^e)^2 + m_e^2} + \sqrt{(p_T^e)^2 + m_\nu^2}
\end{eqnarray}
(which is $\simeq 2 p_T^e$ if $m_e$ and $m_\nu$ are neglected), one can change variables $p_T^e\rightarrow {\cal E}_T$, introducing a trivial Jacobian factor. So in the ${\cal E}_T$ variable
the pole occurs at
\begin{eqnarray}
\label{poleW}
\left.{\cal E}_T\right|_{\rm pole}= E \simeq M_W\ .
\end{eqnarray}
In consequence, the histogram of events in the ${\cal E}_T$ variable should show a peak at ${\cal E}_T=E\sim M_{W}$ followed by an abrupt fall to zero, since ${\cal E}_T\leq E$, . This behavior is slightly softened by the fact that the $W$ has a non-vanishing width.

The previous analysis was done at the CM system, but its extension to the LAB one is straightforward. First of all, since ${\cal E}_T$ is invariant under $z-$boosts, the presence of the pole in ${\cal E}_T$ at the CM energy ($\simeq M_W$) holds in the LAB system assuming there is no net transverse momentum, which is a sensible approximation. Actually, this is a successful strategy to identify $M_W$ in this context. Of course, in practice there can be a net transverse momentum if the collision is not completely along the $z-$line, due mainly to initial state radiation. To incorporate this effect to the analysis note that, in CM frame, the electron and neutrino momenta for {\em the events at the pole} have no longitudinal components, so that $E_T^e = E^e$, $E_T^\nu = E^\nu$, and ${\cal E}_T = E_T^e + E_T^\nu$ coincides with the total energy. These features obviously hold after a boost in a transverse direction. Consequently, in the boosted (LAB) system the pole in the cross section still occurs at 
\begin{eqnarray}
\label{poleWboost}
\left.{\cal E}_T\right|_{\rm pole}= E = \sqrt{E_{CM}^2 + (\vec{p}_T^{\ e}+\vec{p}_T^{\ \nu})^2}\ ,
\end{eqnarray}
where $E_{CM}=M_W$. (Of course the previous expression is invariant under further $z-$boosts.) If desired, the effect of a net transverse momentum of the $W$ can be extracted by defining a ``transverse mass", $M_T$, as
\begin{eqnarray}
\label{eq:MT}
M_{T}^2 &=& {\cal E}_T^2 -(\vec{p}_T^{\ e}+\vec{p}_T^{\ \nu})^2\ = \ 
(E_T^e + E_T^\nu)^2
-(\vec{p}_T^{\ e}+\vec{p}_T^{\ \nu})^2
\nonumber\\
&\simeq& 2\ |{p}_T^{\ e}|\ |{p}^{\ \nu}_T| [1 - \cos\,
  (\phi^e-\phi^\nu)]\ ,
\end{eqnarray}
where in the last expression we have neglected the electron and neutrino
masses, which is allowed in {\em this} context. Note that $M_T$ is invariant
under any boost and satisfies $M_T\leq M_{\rm inv}$, so the cross section does
present a pole at $M_T = M_{W}$. The complication is that the determination of
$\vec{p}_T^{\ e}+\vec{p}_T^{\ \nu}$ is not as clean as one would like (although normally
is a subdominant term). Under the assumption of a perfect longitudinal collision, this
term vanishes and $M_T$ coincides with ${\cal E}_T$.

\subsection{Pair production of SUSY particles at LHC}

Now we consider our scenario of production of supersymmetric particles (in pairs) at the LHC. In most cases the supersymmetric particles produced are squarks and/or gluinos, which decay along jets through diverse channels, plus one lightest supersymmetric particle (LSP, usually a neutralino) for each SUSY particle. Therefore the final state mainly consists of jets (plus possibly some leptons) and two LSPs (plus possibly some neutrinos). This is schematically represented in Fig.~2, where we have not shown the fact that the decays ocurr along two chains, one for each initial supersymmetric particle. Normally the SUSY particles are created close to at-rest in the CM system, since producing them with substantial momentum amounts a high price from the parton distribution functions. Along the paper we will often use this approximation.

\begin{figure}[t]
\begin{center}
\includegraphics[width=0.40\linewidth]{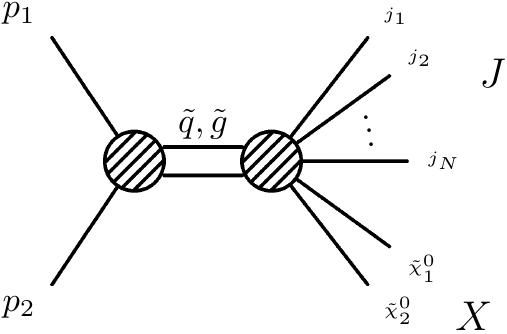}
\end{center}
\caption{Diagram of production and decay of supersymmetric particles at LHC.}
\label{fig:JXdiagram}
\end{figure}

To connect with the strategy of the previous subsection, we can imagine that this process consists of the production of two pseudo-particles, $J$ and $X$, which contain the jets and the two invisible LSPs respectively, as shown in Fig.\ref{fig:JXdiagram}. Hence the momentum and the invariant mass of $J$ ($X$) is the global momentum and invariant mass of the set of jets (invisible particles). In this way, many of the results of the previous subsection can be applied here replacing the $W$-boson and its mass by the two initial supersymmetric particles and their global invariant mass, $M_{\rm inv}$, or equivalently the total energy at CM; and replacing $e\rightarrow J$, $\nu\rightarrow X$. The main difference with the $W$-case is that $J$ and $X$ have masses different
from zero, which are equal to the invariant masses of the visible and invisible systems respectively. These masses, which change from event to event, may be quite
large and cannot be neglected. In addition, $M_{\rm inv}$ does not coincide with the sum of the masses of the initial
supersymmetric particles, since they are not produced exactly at rest,
although this is normally a good approximation.

\begin{figure}[t]
\begin{center}
\includegraphics[width=0.55\linewidth]{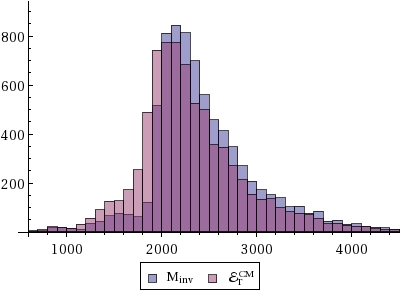}
\end{center}
\caption{${\cal E}_T$ at CM frame (pink) compared with the global invariant mass, $M_{\rm inv}$,  of
the supersymmetric particles initially produced (grey) for the benchmark point SU9 defined in sect.~5.}
\label{fig:ETatCM}
\end{figure}

Now, working in the CM system, we can formally consider a subset of events with
identical structure of jets and invisible particles, differing only in the
$\theta, \phi$ angles at which the pseudo-particle $J$ (and thus $X$) is
produced. Obviously for those processes the phase space contains a factor like
the one given in eq.(\ref{Jacobian}), replacing $e\rightarrow J$,
$\nu\rightarrow X$. And this holds for any jet configuration. So we expect a
similar pole in the ${\cal E}_T-$histogram. This can also be seen by
performing a change of variables $\vec{p}^{\ 1} \rightarrow
\vec{p}^{\ J}=\sum_j \vec{p}^{\ j}$ in the phase space integration of the jet
momenta. One gets
\begin{eqnarray}
\prod_{j=1}^N \frac{d^3 \vec{p}^{\ j}}{(2\pi)^3\, 2E^j} \rightarrow \frac{d^3 \vec{p}^{\ J}}{(2\pi)^3\, 2E^1} 
\prod_{j=2}^N \frac{d^3 \vec{p}^{\ j}}{(2\pi)^3\, 2E^j} 
\end{eqnarray}
and a similar transformation for the momenta of the invisible particles. Then, in
analogy with the $W-$case, we can now consider the differential cross section
with respect to the transverse component of the $J-$momentum, $p_T^J$, obtaining that
eqs.(\ref{dsigma}, \ref{Jacobian}) hold with the $e\rightarrow J$,
$\nu\rightarrow X$ replacements. Thus
\begin{eqnarray}
\label{dsigmaJX}
d \sigma \propto \frac{d \cos \theta}{d p_T^J} d p_T^J\ ,
\end{eqnarray}
where
\begin{eqnarray}
\label{JacobianJX}
\frac{d \cos \theta}{d p_T^J} &=& \frac{-p_T^J |\vec{p}^J| }{\sqrt{|\vec{p}^J|^2-(p_T^J)^2}}
\nonumber\\
&=& \frac{-2E\ p_T^J |\vec{p}^J| }{ \sqrt{E^2- (E_T^J + E_T^X)^2}\sqrt{E^2- (E_T^J - E_T^X)^2}}
\end{eqnarray}
with $E=E^J+E^X$ denoting the total energy in the CM and
\begin{eqnarray}
\label{ETJX}
E_T^J  &=& \sqrt{(p_T^J)^2 + m_J^2}= \sqrt{(E^J)^2-(p_z^J)^2},
\nonumber\\
E_T^X  &=& \sqrt{(p_T^X)^2 + m_X^2}= \sqrt{(E^X)^2-(p_z^X)^2}\ .
\end{eqnarray}
Again, we define a global ``transverse energy" as
\begin{eqnarray}
\label{ETSUSY}
{\cal E}_T = E_T^J + E_T^X\ ,
\end{eqnarray}
so that the differential cross section in the transverse momentum, eqs.(\ref{dsigmaJX}, 
\ref{JacobianJX}), shows a {\em pole} at 
\begin{eqnarray}
\label{poleSUSY}
\left.{\cal E}_T\right|_{\rm pole} = E =M_{\rm inv} \simeq m_1 + m_2\ .
\end{eqnarray}
Here $m_1, m_2$ are the masses of the supersymmetric particles initially
created, which, in the last identity, we have assumed to be produced
approximately at rest in the CM system. This pole is maintained when the cross
section is displayed in the ${\cal E}_T$ variable because the change of
variables $p_T^J\rightarrow {\cal E}_T$ does not introduce any singular
behavior. Note that the position of the pole is always the one given by
eq.(\ref{poleSUSY}) for any subset of events (with any structure of jets and invisible particles) considered. So a global histogram in
the ${\cal E}_T$ variable must show a peak at ${\cal E}_T = E\simeq m_1 +
m_2$. This is illustrated in Fig.~\ref{fig:ETatCM}, which shows the
distribution of ${\cal E}_T$ at CM compared with the invariant mass, $M_{\rm
  inv}$, of the the supersymmetric particles initially produced. The plot
  corresponds to the benchmark point SU9 (defined and discussed in sect. 5)
  where for this particular example we have not included Initial State
  Shower. As one can see, the two histograms are pretty similar, exhibiting a peak at the expected value. This similarity is mainly
  because $J$ and $X$ are normally rather heavy and therefore $m_J$ and $m_X$ are the
  dominant terms in the invariant mass, leading to a small contribution of
  $p_z$, even for events outside the pole.

We have indeed checked the presence of the peak at the expected value (${\cal
  E}_T\simeq m_1 + m_2$) for a large collection of CMSSM and more general MSSM
models, using a PYTHIA simulation. This fact will be used in the next section
to understand the correlation between $M_{\rm susy}$ and $M_{\rm eff}$.
%
%
We postpone the details and discussion of this and other related checks to section 5.

Since ${\cal E}_T$ is invariant under boosts in the $z-$direction, the
previous pole shows up also in the LAB system, provided the net transverse
momentum of the two initial supersymmetric particles is small, which is the usual case. Once more, if
desired, the effect of such net non-vanishing transverse momentum can be
formally extracted by using, instead of ${\cal E}_T$, the transverse mass,
$M_T$, defined as in eq.(\ref{eq:MT}), i.e. $M_{T}^2 = {\cal E}_T^2
-(\vec{p}_T^{J}+\vec{p}_T^{X})^2$. 

%
%


\section{Understanding the correlation $M_{\rm susy}-M_{\rm eff}$}

In this section we show that the empirical correlation between $M_{\rm eff}$ (at the maximum of the histogram of events) and
$M_{\rm susy}$ is an echo of the fact, discussed in the previous subsection, that the cross section has a pole (in practice a maximum) at ${\cal E}_T \simeq m_1 + m_2$.

Since the correlation has to do with the total mass of the two supersymmetric particles initially created, a convenient definition of $M_{\rm susy}$ is
\begin{eqnarray}
\label{eq:Msusy3} 
M_{\rm susy}=\frac{\sum_{a,b}\sigma_{ab}\,
  (m_a+m_b)}{\sum_{a,b}\sigma_{ab}},
\end{eqnarray}
where $a,b$ run over all supersymmetric particles and $\sigma_{ab}$ is the
production cross section of the $\{a,b\}$ pair. Roughly speaking $M_{\rm
  SUSY}$ is the sum of the masses of the two supersymmetric particles in the
dominant channel of production. This definition is exactly twice the
definition of $M_{\rm susy}$ proposed in ref.\cite{Tovey:2000wk} and
approximately twice the HPSSY definition given in eq.(\ref{eq:Msusy1}) if the
dominant channel is squark-squark or gluino-gluino. In the rest of the paper
$M_{\rm susy}$ will always refer to the definition of
eq.(\ref{eq:Msusy3})\footnote{All these definitions may become problematic when
  the the masses of the more frequently produced pairs of superparticles are
  very different. This instance will be discussed in sect. 5.}.

Let us now discuss the definition and meaning of $M_{\rm eff}$. Certainly
$M_{\rm eff}$ has some resemblance with the ${\cal E}_T$ variable defined in eqs.(\ref{ETSUSY}, \ref{ETJX}). In order to deepen in
the connection between both, let us re-write explicit expressions for the two
variables. $M_{\rm eff}$ is defined as
\begin{eqnarray}
\label{eq:meff}
M_{\rm eff}=\sum_{j}
|{p}_T^{\ j}|+ |{p}_T^{\ \mathrm{miss}}|\ \equiv\ M_{\rm eff}^J\ +\ |{p}_T^{\ \mathrm{miss}}|\ ,
\end{eqnarray}
where for the sake of the discussion we have separated the jet contribution to
the effective mass, denoted by $M_{\rm eff}^J$, from the invisible one,
$|{p}_T^{\ \mathrm{miss}}|$. Normally $M_{\rm eff}^J$ is by far the most
important contribution to $M_{\rm eff}$. Note that we have slightly modified
the initial definition of eq.(\ref{eq:meff4}) by extending the sum to all
jets, rather than just the four hardest jets. This seems reasonable and
follows the strategy of ref.\cite{Tovey:2000wk}. A different thing is how to
impose cuts in order to optimally count all the jets coming from the partonic
process, leaving outside all the jets coming from initial state
radiation. This will be addressed in sect. 5.

On the other hand, ${\cal E}_T$, defined in eqs.(\ref{ETSUSY}, \ref{ETJX}), can be re-written in the following way 
\begin{eqnarray}
{\cal E}_T = E^J_T  + E^X_T,  
\label{ETs}
\end{eqnarray}
with
\begin{eqnarray}\nonumber
(E_T^J)^2  &=& (E^J)^2 - (P_z^J)^2 \\ 
&=& \sum_{j=1}^N\left(|\vec{p}_T^{\ j}|^2 + m_j^2\right) 
+ \sum_{i\neq j}^N\sqrt{|\vec{p}_T^{\ i}|^2 + m_i^2} \sqrt{|\vec{p}_T^{\ j}|^2 + m_i^2}\cosh{(y_i-y_j)},
\label{ETJ2}\\\nonumber
(E_T^X)^2  &=& (E^X)^2 - (P_z^X)^2 \\ 
&=& \sum_{\chi_i= 1}^2 \left(|\vec{p}_T^{\ \chi_i}|^2 + m_{\chi_i}^2\right) + 2\sqrt{|\vec{p}_T^{\ \chi_1}|^2 +
  m_{\chi_1}^2}\sqrt{|\vec{p}_T^{\ \chi_2}|^2 + m_{\chi_2}^2} \cosh{(y_{\chi_1}-y_{\chi_2})},
\label{ETX2}
\end{eqnarray}
where $y_j$ and $y_{\chi_{1,2}}$ are the rapidities of the jets and the two neutralinos (or whatever LSPs) respectively.

Comparing the above expressions (\ref{eq:meff})--(\ref{ETX2}) for $M_{\rm eff}$ and ${\cal E}_T$ we see that $M_{\rm eff}^J\leq E^J_T$ and $|p_T^{\rm miss}|\leq E^X_T$, so $M_{\rm eff}\leq {\cal E}_T$. Focusing in $M_{\rm eff}^J$, which is the main contribution to the effective mass, we see that the equality $M_{\rm eff}^J=E_T^J$ is only achieved when the jet masses are negligible, which is often a good approximation, and all jets have the same rapidity, $y_i=y_j$, which is trivially satisfied when there is only one jet, but for two or more jets is never satisfied in practice. Hence, using $M_{\rm eff}^J$ instead of $E^J_T$ makes the peak of the histogram to occur systematically below $m_1+m_2$. 

We can go further by estimating how large is this effect in a typical case. Working in the CM system, the events corresponding to the peak of the histogram have $p_z^J=0$. They satisfy $E_T^J=E^J=\sum_j E^j\simeq \sum_j |\vec{p}^{\ j}|$, where in the last equality we have neglected the jet masses. Consequently, at the peak of the histogram
\begin{eqnarray}
\label{ratio}
\frac{M_{\rm eff}^J}{E_T^J}\simeq\frac{\sum_j|p_T^j|}{ \sum_j |\vec{p}^{\ j}|}\ .
\end{eqnarray}
Usually SUSY searches are done by considering events with a rather large
number of jets, see e.g. refs.\cite{daCosta:2011qk,Aad:2011ib,Aad:2011qa,:2012hm}. Hence, for the sake of the estimate, a reasonable simplification is that (for the events at the peak of the histogram and working in CM) the directions of the various jets are distributed in a more-or-less random way. Then the differential probability that a jet occurs at a particular $\theta$ is $\sin\theta d\theta$. In average $\langle p_T^j\rangle =|\vec{p}^{\ j}|\int d\theta \sin^2\theta=(\pi/4)|\vec{p}^{\ j}|$. Therefore, at the peak of the histogram
\begin{eqnarray}
\label{ratio2}
\frac{<M_{\rm eff}^J>}{E_T^J}\simeq\frac{\pi}{4}=78.5\%\ .
\end{eqnarray}
%
%
This correlation is the main reason for the correlation found between $M_{\rm eff}$ at the peak of the histogram and $\sim 80\%$ $M_{\rm susy}$. 

On the other hand, the $|p_T^{\rm miss}|$ contribution to the effective mass in eq.(\ref{eq:meff}) is also systematically smaller than $E_T^X$, defined in eq.(\ref{ETX2}). As we will see in the next section and Appendix A, for events where the invisible particles are just two neutralinos with momenta larger than their masses, a more fair estimate for $E_T^X$ is $2|p_T^{\rm miss}|$. This means that the invisible contribution to the effective mass is around half the value suitable to get the peak of the histogram at $m_1+m_2$.

We have numerically checked that the peak of the $M_{\rm eff}$ histogram is indeed around 70\%$-$80\% $M_{\rm susy}$; the precise value depends on the model and the cuts used in the analysis. This is illustrated by the statistical survey of CMSSM models presented in Fig. \ref{fig:cmssm} below (light green crosses), which will be discussed in more detail in sect. 4.

\section{Proposal of a new kinematic variable, ${\cal E}_T^{\rm eff}$}

The discussion of the previous section not only allows to understand the correlation between $M_{\rm eff}$ and $M_{\rm susy}$; it also allows to propose an alternative variable which shows an even more robust correlation. As argued above, an histogram in ${\cal E}_T$ shows a maximum near $M_{\rm susy}$; so the idea is simply to devise a new variable which is both measurable and as close as possible to ${\cal E}_T$.

Examining eqs.(\ref{ETs}, \ref{ETJ2}, \ref{ETX2}), we see that $E_T^J$, which is the dominant contribution to ${\cal E}_T$, can in principle be extracted directly from experiment. $E_T^X$ cannot be deduced from the experiment, but is clearly larger than $|p_T^{\rm miss}|$ (the quantity used in the definition of $M_{\rm eff}$), unless the masses of the neutralinos are fairly smaller than their momenta {\em and} the latter have exactly aligned directions, which is unlikely. A much better estimate can be obtained by assuming that the relative directions of the two neutralinos are random in the CM system. This is exact if the the two initial supersymmetric particles are created at rest in CM. Under the further assumption that the 3-momenta of the two neutralinos are similar in magnitude and larger than the neutralino masses, it turns out that, in average, $\langle E_T^X\rangle\simeq 2|p_T^{\rm miss}|$; for more details see Appendix A. Thus our new kinematic variable, say ${\cal E}_T^{\rm eff}$, simply reads
\begin{eqnarray}
\label{ETeff}
{\cal E}_T^{\rm eff} = E_T^J + 2 |{p}_T^{\ \mathrm{miss}}|\ ,
\end{eqnarray}
where $E_T^J$ is given by eq.(\ref{ETJ2}) or eq.(\ref{ETJX}).
In the next section we will test the "performance" of ${\cal E}_T^{\rm eff}$ as a tool to determine $M_{\rm susy}$, and compare it to that of $M_{\rm eff}$, 

To finish this section, let us mention some of the a-priori advantages (and one disadvantage) of ${\cal E}_T^{\rm eff}$ with respect to $M_{\rm eff}$. The most obvious advantage is that 
${\cal E}_T^{\rm eff}$ is much closer to ${\cal E}_T$, and so we expect the peak of the corresponding histogram to be quite close to $M_{\rm susy}$ (which is in fact the case, as we will see). Notice that the definition of ${\cal E}_T^{\rm eff}$ (in particular the jet contribution, $E_T^J$) contains information not only about the transverse components of the jet-momenta but also about the longitudinal ones, thus being more informative than $M_{\rm eff}$. Actually, $M_{\rm eff}$ is proportional to ${\cal E}_T$ only as an average, and the precise proportionality factor changes with analysis-dependent features, such as the number of jets considered for the selected events or the cuts performed on the various kinematical variables. Consequently the peak on the ${\cal E}_T^{\rm eff}-$histogram is likely to indicate more faithfully the value of $M_{\rm susy}$ than the $M_{\rm eff}-$one. Another important advantage of ${\cal E}_T^{\rm eff}$ is that it is robust under the procedure followed to identify the jets. Actually, as it is clear from eq.~(\ref{ETJX}) or (\ref{ETJ2}), for the definition of $E_T^J$ we do not even need to talk about jets: we could perform the sum directly over hadronic final states; or we could consider all the hadronic particles as forming a single jet. The result is the same. On the other hand, a disadvantage of ${\cal E}_T^{\rm eff}$ with respect to $M_{\rm eff}$ is that it relies on a good knowledge of the longitudinal components of the jet momenta. E.g. one might dismiss a jet which really arises from the partonic event because it has large longitudinal component and does not pass the cut in rapidity. This would distort the estimated value of ${\cal E}_T^{\rm eff}$. This happens also for $M_{\rm eff}$, but in the latter case the contribution of such jets is less important, since only the transverse component is counted.

In consequence, ${\cal E}_T^{\rm eff}$ and $M_{\rm eff}$ are complementary variables, rather than competitors; and plotting histograms in both can be useful to cross-check the results, allowing a better and more robust identification of $M_{\rm susy}$. The extension of this procedure to other scenarios of new physics (not necessarily SUSY) is also straightforward.

\section{Testing the efficiency of $M_{\rm eff}$ and ${\cal E}_T^{\rm eff}$ in assorted SUSY models}

We will discuss now the comparative behaviour of $M_{\rm eff}$ and ${\cal
  E}_T^{\rm eff}$, and other kinematic variables, in the context of different
MSSM models. We will simulate LHC signals at $14$ TeV center-of-mass energy
using SOFTSUSY \cite{Allanach:2001kg} and PYTHIA version $6.419$
\cite{Sjostrand:2006za}. Focusing on events with multijets + missing
transverse momentum, and applying the following cuts
\begin{enumerate}
\item At least three jets with $p_T>50$ GeV.
\item The hardest jet with $p_T>100$ GeV and $|\eta|<1.7$.
\item $p_T^{\mathrm{miss}}>100$ GeV.
\item $\Delta\phi(\mathrm{jet}_1-p_T^{\mathrm{miss}})>$ 0.2,
  $\Delta\phi(\mathrm{jet}_2-p_T^{\mathrm{miss}})>$0.2,
  $\Delta\phi(\mathrm{jet}_3-p_T^{\mathrm{miss}})>$0.2.
\end{enumerate}
For the evaluation of the various kinematic variables we will count all the
jets (with $p_T>50$ GeV, as mentioned above)\footnote{We have checked (using a few
  SUSY points) the
  stability of ${\cal E}_T^{\rm eff}$ when applying a cut in $\eta$. Once one requires $p_T>50$ GeV, the impact of such cut is very small. The histograms are very similar whether one uses a $|\eta|\lesssim 3$ cut or not cut at all.}. For the construction of the jets we will use FASTJET
\cite{Cacciari:2011ma}, with the antikt algorithm, $E$ scheme and $R=0.4$.

Let us start by illustrating the behavior of the various kinematic variables
by considering a typical SUSY model, namely the benchmark point $SU9$, defined
in ref. \cite{Aad:2009wy} and specified by the following values of the
supersymmetric parameters:
\begin{eqnarray}
\label{SU9}
m=300\ {\rm GeV},\;\;\;M_{1/2}=425\ {\rm
  GeV},\;\;\;A=20\ ,\;\;\;\tan\beta=20,\;\;\;\mu>0 .
\end{eqnarray}
where $m$, $M_{1/2}$ and $A$ are the universal scalar mass, gaugino
mass and trilinear scalar coupling (all quantities defined at the $M_X$ scale) respectively; $\tan \beta \equiv v_2/v_1$ is the
ratio between the vev's of the two Higgs doublets, $H_1, H_2$; and
$\mu$ is the usual Higgs mass term in the superpotential, defined also at $M_X$. The corresponding values of the squark mass (first two generations) and the gluino mass are $m_{\tilde q}=920$ GeV, $M_{\tilde g}=994$ GeV. The corresponding value of $M_{\rm susy}$, defined according to eq.(\ref{eq:Msusy3}), is 
\begin{eqnarray}
\label{MSUSYSU9}
M_{\rm susy}= 1898\ {\rm GeV} \ .
\end{eqnarray}
\begin{figure}[t]
\begin{center}
\includegraphics[width=0.48\linewidth]{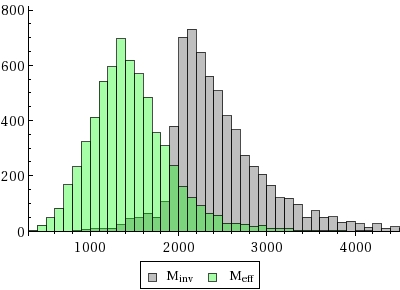}\hspace{0.2cm}
\includegraphics[width=0.48\linewidth]{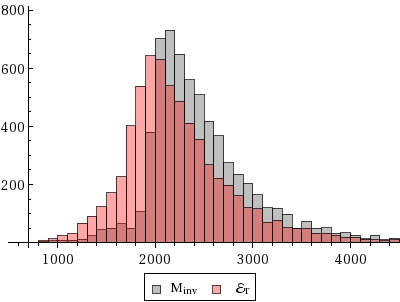}\\ \vspace{0.7cm}
\includegraphics[width=0.48\linewidth]{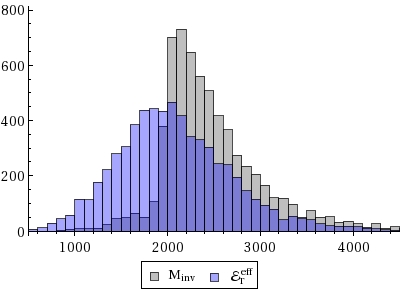}\hspace{0.2cm}
\includegraphics[width=0.48\linewidth]{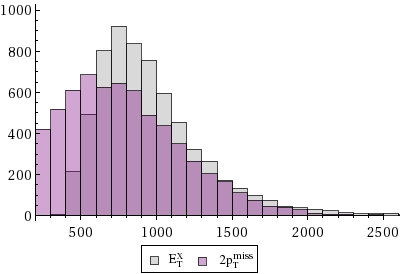}
\end{center}
\caption{$M_{\rm inv}$ and $M_{\rm eff}$ (top-left panel), $M_{\rm inv}$ and
  ${\cal E}_T$ (top-right panel), $M_{\rm inv}$ and ${\cal E}_T^{\rm eff}$
  (bottom-left panel) and $2 |{p}_T^{\ \mathrm{miss}}|$ and $E_T^X$
  (bottom-right panel) for the SU9 benchmark point.}
\label{fig:SU9}
\end{figure}  
Figure \ref{fig:SU9} (top-left panel) shows the histogram distributions of the
invariant mass, $M_{\rm inv}=E_{CM}$, and the histogram distribution of
$M_{\rm eff}$. Clearly the $M_{\rm inv}$ distribution shows a sharp starting
point, almost coinciding with the maximum, at $M_{\rm inv}\simeq M_{\rm
  SUSY}$. Although the peak is quite sharp, the width of the distribution
indicates to what extent the initial supersymmetric particles are created with
non-vanishing momenta at CM. Of course $M_{\rm inv}$ is not directly
measurable, so this histogram cannot be built in practice. On the other hand,
the $M_{\rm eff}-$histogram shows indeed a maximum correlated with $M_{\rm
  SUSY}$, namely $M_{\rm eff}\simeq 70\%\ M_{\rm susy}$ in this case.  Figure
\ref{fig:SU9} (top-right panel) shows the distribution of the ${\cal E}_T$
variable compared with the $M_{\rm inv}-$one. As expected from the discussion
of sect. 2, the maximum of the two histograms are indeed very close.
Comparing this with the ${\cal E}_T^{\rm CM}$ histogram, figure
  \ref{fig:ETatCM}, one can see that ${\cal E}_T$ is shifted to the left
  with respect to ${\cal E}_T^{\rm CM}$. Note that now we are doing a more
  realistic simulation, including Initial State Shower (ISS), so, in order to
  avoid contamination from these ISS, one has to require jets with high $p_T$
  (as described above). Hence, the main reason for this shift is that
  by including those cuts we are loosing some of the final particles coming from
  supersymmetric decays, which leads to an underestimate of the visible part; but still the maximum is where was expected.
Fig \ref{fig:SU9} (bottom-left panel) is a similar plot, but using the
measurable variable, ${\cal E}_T^{\rm eff}$, defined in eq.(\ref{ETeff}). We
see that the ${\cal E}_T^{\rm eff}-$histogram maintains the peak close to
$M_{\rm susy}$, though is somewhat less sharp than in the ${\cal
  E}_T-$histogram. This is due to the fact that the invisible contribution to
${\cal E}_T^{\rm eff}$ in eq.(\ref{ETeff}), i.e. $2
|{p}_T^{\ \mathrm{miss}}|$, is an average of $E_T^X$, i.e. the actual
contribution, to ${\cal E}_T$; see eqs. (\ref{ETs}, \ref{ETX2}) and Appendix
A. The goodness of such average (quite satisfactory in this case) can be
appreciated in Figure \ref{fig:SU9} (bottom-right panel), where these two
quantities are plotted.

Let us now explore how good is the behavior of $M_{\rm eff}$ and ${\cal E}_T^{\rm eff}$ in general SUSY models. We start considering a sample of 500 random CMSSM points requiring dominant $\tilde{q}\tilde{q},\,\tilde{g}\tilde{q},\, \tilde{g}\tilde{g}$ production, which is the usual case. Recall that the CMSSM is defined by the values of
\begin{displaymath}
\label{CMSSM}
m,\ M_{1/2},\ A,\ \tan\beta,\ {\rm sign }\mu\; .
\end{displaymath}
(Incidentally, the benchmark point $SU9$, defined at eq.(\ref{SU9}) was a particular CMSSM model.) For all the models we have performed the simulation of the proton proton collisions using the specifications presented at the beginning of this subsection.

\begin{figure}[t]
\centering 
\includegraphics[width=0.65\linewidth]{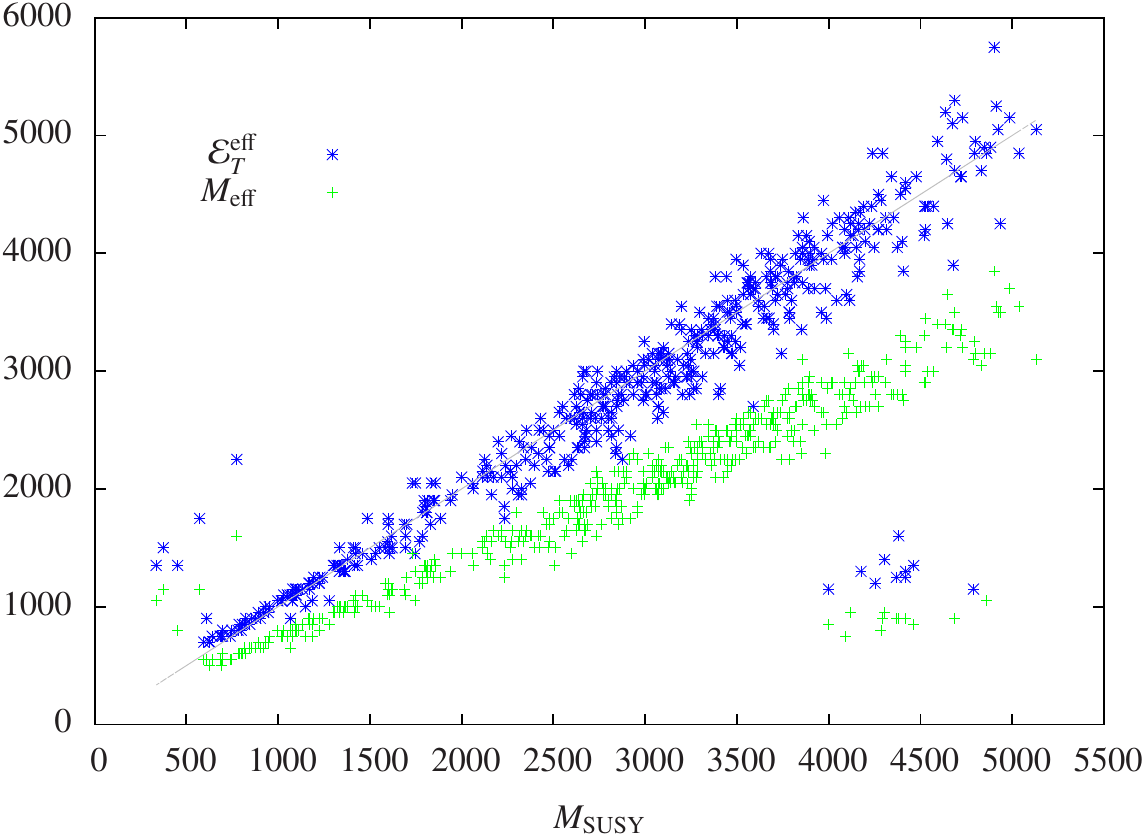}
\caption{$M_{\rm susy}$ versus ${\cal E}_T^{\rm eff}$ (blue stars) and $M_{\rm eff}$ (light green
  crosses) for the CMSSM.}
\label{fig:cmssm}
\end{figure}

Figure \ref{fig:cmssm} shows the values of ${\cal E}_T^{\rm eff}$ (blue stars) and 
and $M_{\rm eff}$ (light green crosses), which maximize their corresponding
histograms, versus $M_{\rm susy}$ for those 500 CMSSM models.

As expected from the previous discussion, there is an remarkable correlation between ${\cal E}_T^{\rm eff}$ at the maximum of the histogram and $M_{\rm susy}$; namely ${\cal E}_T^{\rm eff}$ $\simeq$ $M_{\rm susy}$. The correlation is also good for the effective mass: $M_{\rm eff}$ (at the histogram maximum)$\simeq$ $70\%\ M_{\rm susy}$, confirming the empirical observations of previous literature \cite{Hinchliffe:1996iu,Tovey:2000wk}. 

Notice that, for large values of $M_{\rm susy}$, there are few points which do not correlate well. This is because at large energies the production of charginos and neutralinos may become competitive. In that case there appear two separate peaks in the histograms, which makes the definition of $M_{\rm susy}$ fairly contrived.

Let us now extend the previous study to more general MSSM models, allowing non-universal soft parameters at the $M_X$ scale. For that goal, we extend the previous parameter space, eq.(\ref{CMSSM}), to a 15-dimensional parameter space, defined by
\begin{displaymath}
M_1,M_2,M_3;\ A_t,A_b,A_\tau; \ m_{H_u,H_d};
\ m_{\tilde{q}_L,\tilde{q}_R},m_{\tilde{t}_L,\tilde{b}_L},m_{\tilde{t}_R},m_{\tilde{b}_R};
\ m_{\tilde{l}_L,\tilde{l}_R},m_{\tilde{\tau}_L,\tilde{\nu_\tau}_L},m_{\tilde{\tau}_R};
\ \tan\beta .
\end{displaymath}
Here $M_a$ are the (non-universal) gaugino masses; and $A_i$, $m_i$ are the non-universal trilinear scalar couplings and scalar masses ($i$ denotes flavour species). $\tilde{q}$, $\tilde{l}$ denote squarks and sleptons of the first two generations. 
As for the CMSSM case, we have studied 500 models chosen at random in this parameter space.

\begin{figure}[t]
\centering 
\includegraphics[width=0.48\linewidth]{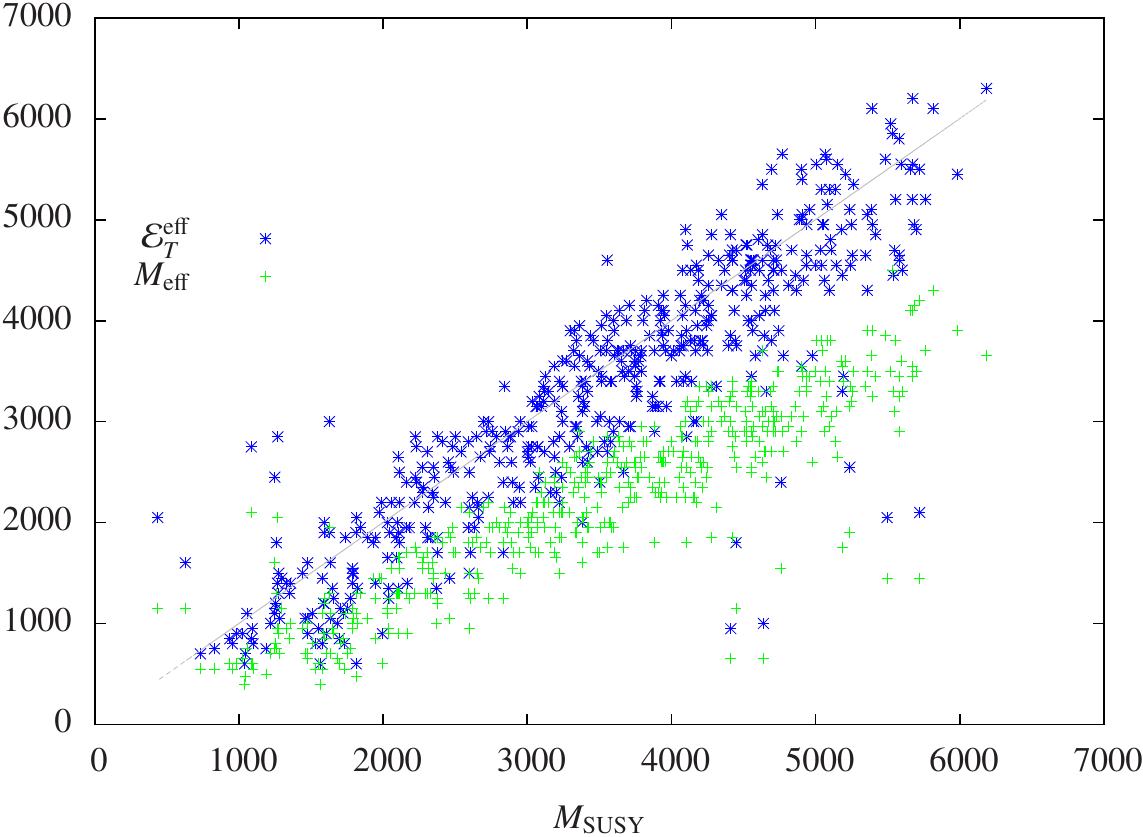} \,\,\,\,\, 
\includegraphics[width=0.48\linewidth]{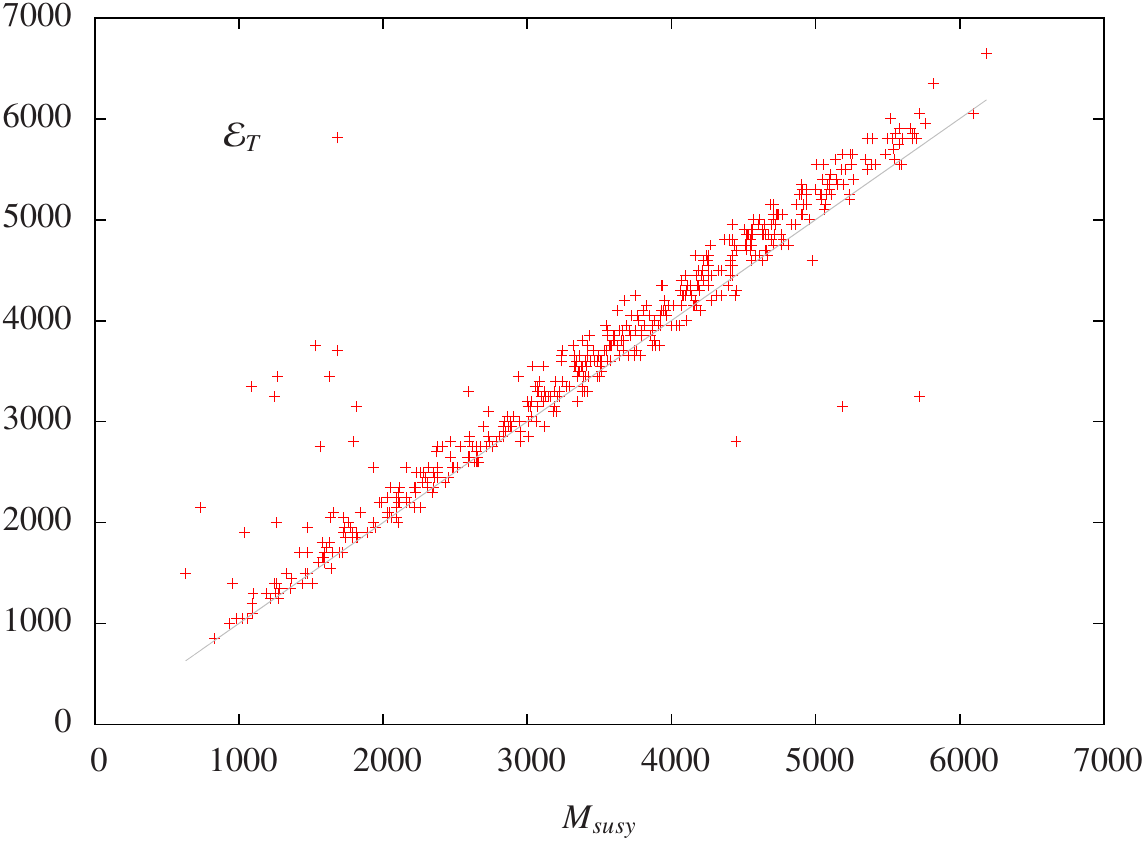}
\caption{Left panel shows $M_{\rm susy}$ versus ${\cal E}_T^{\rm eff}$ (blue
  stars) and $M_{\rm eff}$ (light green crosses). Right panel shows $M_{\rm
    SUSY}$ versus ${\cal E}_T$ (red crosses). Both for the MSSM.}
\label{fig:mssm}
\end{figure}

Figure \ref{fig:mssm} (left panel) is as Figure \ref{fig:cmssm} but within this more general MSSM scenario. In this case the correlation is not as good
as for the CMSSM, something that was empirically noticed in ref.\cite{Tovey:2000wk} for the $M_{\rm eff}$ variable. We have checked that the reason is that now the neutralino masses can
be much larger than in typical CMSSM models. Then, since in the definition of both $M_{\rm eff}$ and ${\cal E}_T^{\rm eff}$ one is neglecting the neutralino masses, one is missing a potentially important piece. Furthermore, in some cases many neutrinos can be produced. 
This makes the average missing transverse energy, $E_T^X$, larger than the average $\simeq 2 |p_T^{\rm miss}|$ used in the definition of ${\cal E}_T^{\rm eff}$ (and of course larger than the $|p_T^{\rm miss}|$ piece entering the definition of $M_{\rm eff}$). This can be checked by plotting the true ${\cal E}_T$ variable (which cannot be directly measured). This has been done in the right panel of figure \ref{fig:mssm}. Remarkably, here the correlation with $M_{\rm susy}$ is nicely maintained, indicating that the spreading exhibited by ${\cal E}_T^{\rm eff}$  and $M_{\rm eff}$ is indeed due to an underestimate of the invisible contribution.

The previous discussion suggests that a better correlation could be found by refining the estimate $\langle E_T^X\rangle\simeq 2|p_T^{\rm miss}|$ used above, by taking into account the finite size of neutralino masses. As discussed at the end of the Appendix, a more refined estimate of the invisible transverse energy for the events at the pole is obtained by retaining the dominant terms in $m_\chi^2$, and reads $\langle E_T^X\rangle \simeq \sqrt{4(p_T^X)^2 + 4 m_\chi^2}\simeq 2p_T^X + 2m_\chi^2/E_T^X$. Using the fact that typically $E_T^X$ contributes less than half to ${\cal E}_T$, we get a conservative correction $\sim 4 m_\chi^2/M_{\rm susy}$ to ${\cal E}_T^{\rm eff}$ for the events at the pole. Alternative, we can keep the definition of ${\cal E}_T^{\rm eff}$ given at eq.~(\ref{ETeff}). Then the value of ${\cal E}_T^{\rm eff}$ at the maximum of the histogram must be around
\begin{eqnarray}
\label{MSUSYcorr}
M_{\rm susy} - 4 \frac{m_\chi^2}{M_{\rm susy}}\ ,
\end{eqnarray}
rather than $M_{\rm susy}$. Incidentally, a similar correction to $M_{\rm susy}$ with the same functional dependence but half the value, was proposed in ref.\cite{Tovey:2000wk}.

\begin{figure}[t]
\centering 
\includegraphics[width=0.65\linewidth]{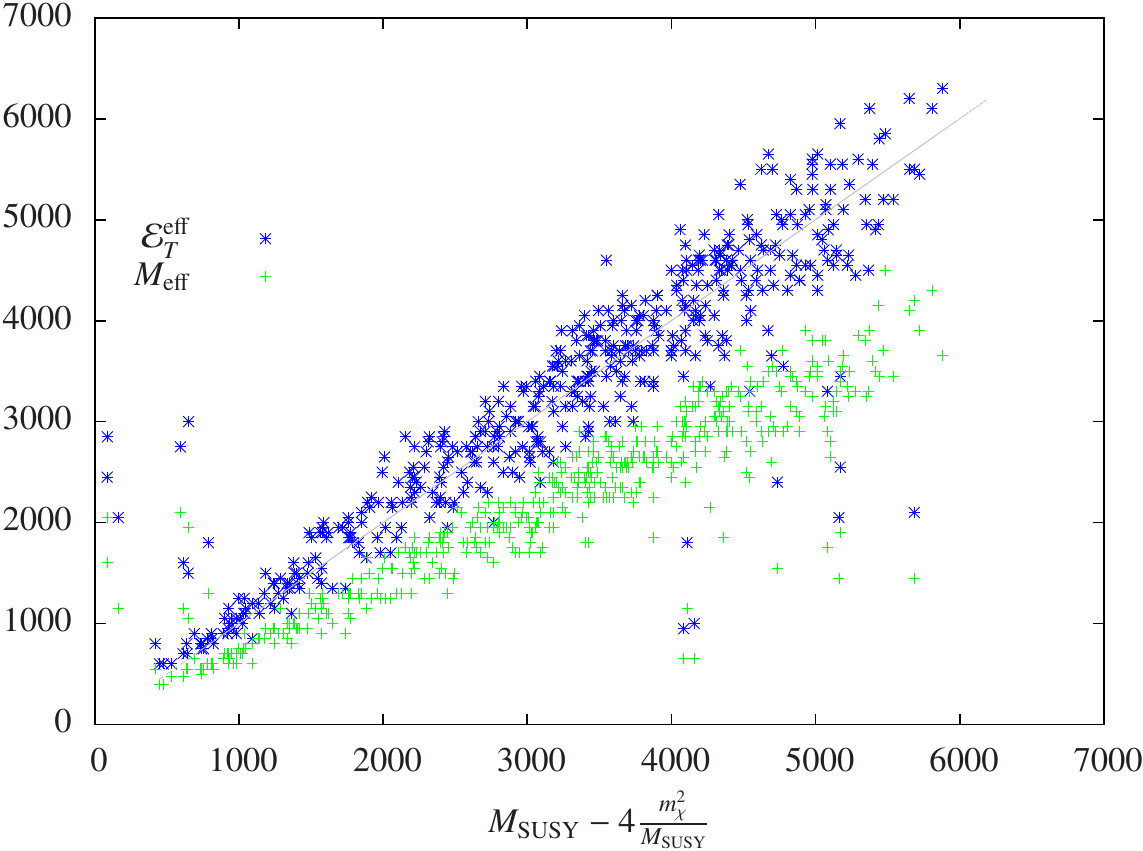}
\caption{$M_{\rm susy} - 4 \frac{m_\chi^2}{M_{\rm susy}}$ versus ${\cal E}_T^{\rm eff}$ (blue
  stars) and $M_{\rm eff}$ (light green crosses) for the MSSM.}
\label{fig:mssm2}
\end{figure}

Fig.~\ref{fig:mssm2} shows $M_{\rm susy} - 4 \frac{m_\chi^2}{M_{\rm susy}}$ versus ${\cal E}_T^{\rm eff}$ for the same 500 general MSSM models displayed in Fig.~6. Clearly the correlation is now much better, comparable to that found for CMSSM models (Fig.~\ref{fig:mssm}). The correction (\ref{MSUSYcorr}) can also be applied to CMSSM models, but in that case the improvement is less significant since $m_\chi$ is normally much smaller than $M_{\rm susy}$.

\section{Conclusions}

There is an
intense activity to explore assorted strategies for the search of
new physics at the LHC. A paradigmatic case is supersymmetry (SUSY).
Under the assumption that superparticles are light
enough to be produced and detected at LHC, one can expect a typical
signal of multijets and large missing energy. However, the analysis is quite complicated and the correct
identification and reconstruction of the properties of the final
states is quite tough. Besides, the separation of the decay chains is normally not
possible. In this context the use of appropriate kinematic variables is instrumental to maximize the signal of new physics from the Standard Model background.

A handy and extensively used kinematic variable, e.g. in ATLAS analyses, is the Effective Mass variable, defined as 
\begin{eqnarray}
\label{eq:meff2}
M_{\rm eff}=\sum_{j}
|{p}_T^{\ j}|+ |{p}_T^{\ \mathrm{miss}}|\ \equiv\ M_{\rm eff}^J\ +\ |{p}_T^{\ \mathrm{miss}}|\ .
\end{eqnarray}
It was empirically found in refs.\cite{Hinchliffe:1996iu,Tovey:2000wk} that for the Constrained Minimal Supersymmetric Standard Model (CMSSM) there is an outstanding correlation between the value of $M_{\rm eff}$ at which the histogram of events has a maximum and the typical supersymmetric masses. Namely, 
\begin{eqnarray}
\label{corr}
\left. M_{\rm eff}\right|_{\rm max}\simeq 80\%\ M_{\rm susy}\ ,
\end{eqnarray}
where $M_{{\rm susy}}\sim m_1+m_2$, i.e. the sum of the masses of the supersymmetric particles initially created (usually squark-squark, gluino-gluino or squark-gluino). However, for some points in the CMSSM, and especially in the general MSSM, the correlation fails.

In this paper we have found an explanation for the above correlation (\ref{corr}). We have argued that $M_{\rm eff}$ is typically $80\%$ of a kinematic variable, ${\cal E}_T$, defined as the sum of the transverse energies of the visible and the invisible parts of the decay products of the initial supersymmetric particles; for more details see eqs.(\ref{ETJX}) and (\ref{ETJ2}, \ref{ETX2}). Then it can be shown that, at CM, the cross section has a maximum at ${\cal E}_T=E_{\rm CM}\sim m_1+m_2$. This relation is invariant under longitudinal boosts and thus is kept as a good approximation in the LAB frame. 

This understanding has allowed us to propose a new kinematic variable, the effective transverse energy ${\cal E}_T^{\rm eff}$, which is measurable and as close as possible to ${\cal E}_T$. It reads
\begin{eqnarray}
\label{ETeff2}
{\cal E}_T^{\rm eff} = E_T^J + 2 |{p}_T^{\ \mathrm{miss}}|.
\end{eqnarray}
where $(E_T^J)^2=(E^J)^2-(p_z^J)^2$ is the hadronic transverse energy [for more details see eqs.(\ref{ETJX}), (\ref{ETJ2})]. ${\cal E}_T^{\rm eff}$ is a kinematic variable alternative to $M_{\rm eff}$, 
which shows an even better and more direct correlation with the supersymmetric mass
\begin{eqnarray}
\label{corr2}
\left. {\cal E}_T^{\rm eff} \right|_{\rm max}\simeq M_{\rm susy}\ .
\end{eqnarray}
Besides, ${\cal E}_T^{\rm eff}$ has other advantages, like being more robust under the procedure followed to identify the jets (actually, it is completely independent of it). On the other side, the determination of ${\cal E}_T^{\rm eff}$ relies (more than for $M_{\rm eff}$) on a good knowledge of the longitudinal components of the jet momenta. 
In consequence, ${\cal E}_T^{\rm eff}$ and $M_{\rm eff}$ are complementary variables, rather than competitors; and plotting histograms in both can be useful to cross-check the results, allowing a better and more robust identification of $M_{\rm susy}$. The extension of this procedure to other scenarios of new physics (not necessarily SUSY) is also straightforward.

We have shown these features in the context of the CMSSM or more general MSSM models by examining the above correlations in a large number of points in the parameter space. The results show that ${\cal E}_T^{\rm eff}$ is a simple and valuable kinematic variable to determine the characteristic supersymmetric masses.

Our analysis also shows why the above correlations fail for some models. This is because at large energies the production of charginos and neutralinos may become competitive. In those cases there appear two separate peaks in the histograms, which makes the previous analysis too simple. In addition, for general MSSM models (departing from the conventional CMSSM) the neutralino masses can
be much larger than in typical CMSSM models. Furthermore, in some cases many
neutrinos can be produced. In those cases, the term representing the missing
transverse energy in (\ref{eq:meff2}) and even in (\ref{ETeff2})
underestimates the actual invisible contribution and a more refined estimate of it must be used. Then, retaining the dominant contributions in the neutralino mass, the right-hand side of eq.~(\ref{corr2}) gets shifted as $M_{\rm susy} \rightarrow  M_{\rm susy}- 4 {m_\chi^2}/{M_{\rm susy}}$, significantly improving the correlation.

\section{Appendix}

\begin{figure}[t]
\centering 
\includegraphics[width=0.58\linewidth]{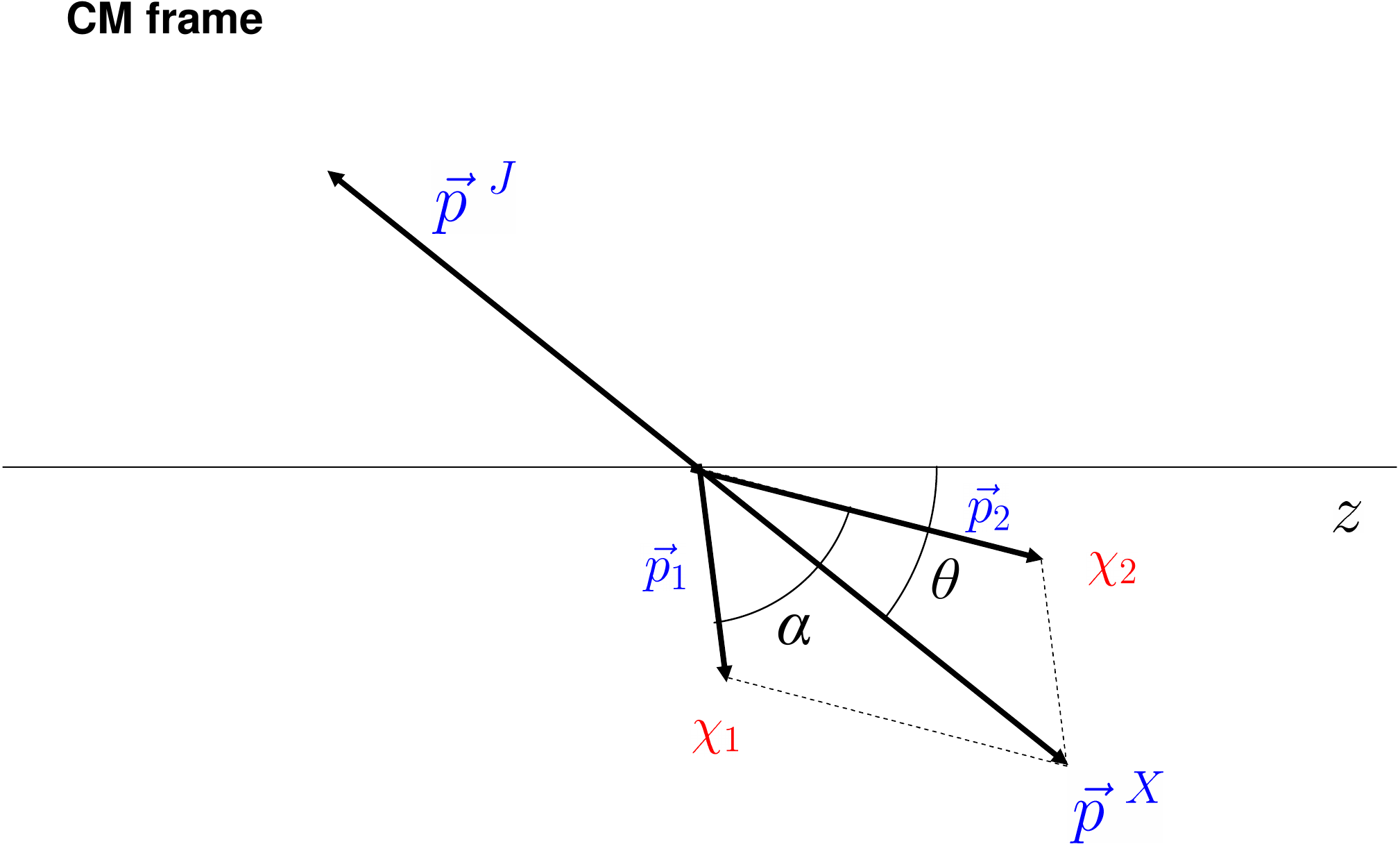}
\caption{}
\label{fig:angulos}
\end{figure}

In this appendix we evaluate the average value of the missing transverse energy, defined in eq.(\ref{ETJX}),
\begin{eqnarray}
\label{ETXAp}
(E_T^X)^2  = (p_T^X)^2 + m_X^2= (E^X)^2-(p_z^X)^2\ .
\end{eqnarray}
Here $p_T^X$, $p_z^X$ are the transverse and longitudinal components of the total invisible 4-momentum
\begin{eqnarray}
\label{pp1p2}
p^X = p_1 + p_2\ ,
\end{eqnarray}
where $p_1, p_2$ are the two 4-momenta of the two final-state neutralinos, $\chi_1, \chi_2$. From now on we will work in the CM reference frame, see Fig.\ref{fig:angulos}. 
As represented in Fig.\ref{fig:angulos}, $\theta$ denotes the angle of $\vec{p}^{\ X}$ with the longitudinal axis, $z$; and $\alpha$ the angle between $\vec{p}_1$ and $\vec{p}_2$. Let us now write the missing energy, $E_T^X$, using the second equality in eq.(\ref{ETXAp}):
\begin{eqnarray}
\label{}
\hspace{-7mm}(E_T^X)^2  &=& \left(\sqrt{\vec{p}_1^{\ 2}+m_\chi^2} + \sqrt{\vec{p}_2^{\ 2}+m_\chi^2}\right)^2 -(\vec{p}^{\ X})^2\cos^2\theta
\nonumber\\
&=& 2m_\chi^2 + 2\sqrt{(\vec{p}_1^{\ 2}+m_\chi^2)(\vec{p}_2^{\ 2}+m_\chi^2)} + \left( \vec{p}_1^{\ 2} +\vec{p}_2^{\ 2} \right)\sin^2\theta - 2 |\vec{p}_1| |\vec{p}_2| \cos\alpha\cos^2\theta .
\end{eqnarray}
On the other hand, the invisible transverse momentum, $p_T^X$, is given by
\begin{eqnarray}
\label{}
(p_T^X)^2 = \left( \vec{p}_1^{\ 2} +\vec{p}_2^{\ 2} \right)\sin^2\theta + 2 |\vec{p}_1| |\vec{p}_2| \cos\alpha\sin^2\theta\ .
\end{eqnarray}

Hence,
\begin{eqnarray}
\label{ETX2app}
(E_T^X)^2 = (p_T^X)^2 + 2m_\chi^2 + 2\sqrt{(\vec{p}_1^{\ 2}+m_\chi^2)(\vec{p}_2^{\ 2}+m_\chi^2)} - 2 |\vec{p}_1| |\vec{p}_2| \cos\alpha\ .
\end{eqnarray}
This expression admits further simplifications under some reasonable assumptions. Assuming $m_\chi^2\ll (\vec{p}_i)^2 $, we have
\begin{eqnarray}
\label{}
\hspace{-4mm}(E_T^X)^2 \simeq (p_T^X)^2 + 2 |\vec{p}_1| |\vec{p}_2| (1-\cos\alpha) = (p_T^X)^2 \left[1+
\frac{2 |\vec{p}_1| |\vec{p}_2| (1-\cos\alpha) }{\sin^2\theta\left(\vec{p}_1^{\ 2} +\vec{p}_2^{\ 2} + 2 |\vec{p}_1| |\vec{p}_2| \cos\alpha\right) }
\right].
\end{eqnarray}
Finally, we can assume that typically $|\vec{p}_1| \sim |\vec{p}_2| $. Then
\begin{eqnarray}
\label{}
(E_T^X)^2 \simeq (p_T^X)^2 \left[1+
\frac{1-\cos\alpha }{\sin^2\theta (1+\cos\alpha)} 
\right]\ .
\end{eqnarray}
Now, we can single out the previous expression for the events at the pole, i.e. with $\theta=\pi/2$. Notice also that, assuming that the initial supersymmetric particles are approximately at rest, it turns out that the relative directions of $\vec{p}_1 $ and $\vec{p}_2 $ are random, so the differential probability for a particular value of the angle $\alpha$ is ${\cal P}(\alpha)= \frac{1}{2}\sin\alpha\ d\alpha$. In conclusion
\begin{eqnarray}
\label{ETXfinal}
\langle E_T^X\rangle_{\rm pole}\ \simeq\ \frac{1}{2} p_T^X\ \int_0^\pi d\alpha\ \sin\alpha \left[1+
\frac{1-\cos\alpha }{1+\cos\alpha} \right]^{1/2}\ =\ 2\ p_T^X\ ,
\end{eqnarray}
which is the estimate used of $E_T^X$ throughout most of the
paper. The goodness of this approximation is illustrated in Fig.\ref{fig:SU9}
(bottom-right panel) for a typical example.

A more refined estimate can be obtained by keeping the dominant terms in $m_\chi^2$ in 
eq.~(\ref{ETX2app}). Then, eq.~(\ref{ETXfinal}) gets slightly modified:
\begin{eqnarray}
\label{ETXfinal2}
\langle E_T^X\rangle_{\rm pole} \simeq \sqrt{4(p_T^X)^2 + 4 m_\chi^2}\ .
\end{eqnarray}

\section*{Acknowledgements}  

We are extremely grateful to Juan Terr\'on for illuminating discussions and
his help with statistical tools. Likewise we thank Agust\'{i}n Sabio, Roberto
Ruiz de Austri, Gianfranco Bertone and Paul de Jong for interesting
discussions and suggestions.

This work has been partially supported by the MICINN, Spain, under contract
FPA2010-17747; Consolider-Ingenio PAU CSD2007-00060, CPAN CSD2007-00042. We
thank as well the Comunidad de Madrid through Proyecto HEPHACOS S2009/ESP-1473
and the European Commission under contract PITN-GA-2009-237920. M. E. Cabrera
acknowledges the financial support of the CSIC through a predoctoral research
grant (JAEPre 07 00020); and the ERC ``WIMPs Kairos - the moment of truth for
wimp dark matter'' (P.I. Gianfranco Bertone).

\bibliography{references} 
\bibliographystyle{JHEP}

\end{document}